\documentclass[11pt]{article}
\usepackage{authblk}
\usepackage{natbib}
\usepackage{latexsym}
\usepackage{graphics}
\usepackage{amsfonts}
\usepackage{enumerate}
\usepackage{graphicx}
\usepackage{verbatim}
\usepackage{subfig}
\usepackage{caption}
\usepackage{footnote}
\usepackage{lscape}
\usepackage{hhline}
\usepackage{array}
\usepackage{graphicx}
\usepackage{setspace}
\usepackage{amssymb}
\usepackage{amsmath}
\usepackage{amsfonts}
\usepackage{times}
\usepackage{color}
\usepackage{amsthm}
\usepackage{indentfirst}
\usepackage{epsfig}
\usepackage{lscape,ctable,rotating}
\usepackage{enumitem}
\setlist{nolistsep}
\usepackage[left=1in,top=1in,right=1in,foot=1in]{geometry}

\theoremstyle{definition}

\theoremstyle{remark}

\theoremstyle{plain}

\newcommand{\dt}{{\varDelta t}}

\title{Another Look at the Ho-Lee Bond Option Pricing Model}
\author[1]{Young Shin Kim}
\author[1]{Stoyan Stoyanov}
\author[2]{Svetlozar Rachev}
\author[3,*]{Frank J. Fabozzi}
\affil[1]{College of Business, Stony Brook University}
\affil[2]{Department of Mathematics \& Statistics, Texas Tech University}
\affil[3]{EDHEC Business School}
\affil[*]{Contact author: 858 Tower View Circle, New Hope, PA 18938 USA~ e-mail:fabozzi321@aol.com}
\date{}

\begin{document}
\maketitle
\begin{abstract}
\begin{normalsize}
In this paper, we extend the classical Ho-Lee binomial term structure model to the case of time-dependent parameters and, as a result, resolve a drawback associated with the model. This is achieved with the introduction of a more flexible no-arbitrage condition in contrast to the one assumed in the Ho-Lee model.
\end{normalsize}~
\end{abstract} 
\section{Introduction}
The most commonly used model for pricing bond options, swaptions, caps, and floors is the model developed by \cite{HoLee:1986}. Referred to as the Ho-Lee model, this model was the first arbitrage-free model for pricing interest rate derivatives. In this paper we extend the original Ho-Lee binomial model (\cite{HoLee:1986}) that takes into account the term structure of interest rates for pricing bond options but allows for time-dependent model parameters. By doing so, we resolve a drawback of the Ho-Lee model pointed out by \cite{BlissRonn:1989}: In the Ho-Lee model the one-period short rates at time $t\ge0$ becomes unbounded if a sufficiently long sequence of downturns in the binomial tree is observed. Our binomial model is free of this drawback of the Ho-Lee model.

\section{Ho-Lee Binomial Interest Rate Model with Time-Dependent Parameters}
Our objective is to extend the original Ho-Lee model to the case when the model parameters are dependent, and as a result resolve a limitation of the original Ho-Lee model.

Consider a space $\Omega$ and a discrete-trading time $\Pi(\dt) = \{t=k\dt:k=0, 1,2,\cdots\}$.
Let $(\varepsilon(t))_{t\in\Pi(\dt)}$ be a discrete-stochastic process such that $\varepsilon(0) = 0$ and $\varepsilon(t)$, $t>0$ is Bernoulli distributed with probability $P(\varepsilon(t)=1)=p(t)$, and suppose $\varepsilon(t)$ and $\varepsilon(s)$ are independent if $t\neq s$. For $t=m\dt\in\Pi(\dt)$, we define an $m$-dimensional random vector $E(t) = (\varepsilon(\dt),\varepsilon(2\dt),\cdots,\varepsilon(m\dt))$.
In the Ho-Lee model\footnote{See \citeauthor{HoLee:1986} (\citeyear{HoLee:1986}, \citeyear{HoLee:2009}), \cite{BlissRonn:1989}, \cite{Sommer:1996}, \cite{Prigen:2003}, and \cite{Akahori_et_al:2006}.}  the zero-coupon prices can be represented as a binomial pricing lattice in $\Pi(dt)$. The zero price dynamic at time $t\in\Pi(\dt)$ with maturity $T \in\Pi(\dt)$, $t\le T$, is given by
\begin{equation}\label{eq BinTree}
B(t,T|E(t)) = \frac{B(t-\dt,T|E(t-\dt))}{B(t-\dt,t|E(t-\dt))}\mathbb{H}(\varepsilon(t),t,T) 
\end{equation}
where 
\begin{equation*}
\mathbb{H}(\varepsilon(t),t,T)=
\begin{cases}
U(t,T) & \text{ if }  \varepsilon(t) = 1 \\
D(t,T) & \text{ if }  \varepsilon(t) = 0
\end{cases}.
\end{equation*}

In the original Ho-Lee model, $p(t)=p$ for some constant $p\in(0,1)$. There have been several extensions of the Ho-Lee model\footnote{See the review of Ho-Lee type models in \cite{ArtamanovaLeipus:2005}.}, but in all extensions the probabilities for ``up'' and ``down'' movements are constant.  We consider a time-varying probability for up movements in order to obtain a more flexible model for the price dynamics of the zero-coupon bond.

The binomial tree defined by \eqref{eq BinTree} should be a recombined tree, implying that \footnote{See the Appendix for more details.},
\begin{equation}\label{recombine}
\frac{U(t,T)}{D(t,T)} = \frac{U(t, t+\dt)U(t+\dt,T)}{D(t, t+\dt)D(t+\dt,T)}.
\end{equation}
Set $H(s,t):=\frac{U(s,t)}{D(s,t)}$. Then, \eqref{recombine} equals $H(t, T) = H(t, t+\dt) H(t+\dt, T)$.
Let $h(s,t) = \ln H(s,t)$. Since $t, T \in \Pi(\dt)$ and $t\le T$, there is positive integer $m$ and $N$ such that $t=m\dt$ and $T=N\dt$, we obtain
\[
h(m\dt, N\dt) = h(m\dt, (m+1)\dt)+h((m+1)\dt, N\dt).
\]
Putting $n=N-m$ (i.e., $m=N-n$), we have
\begin{equation}\label{eq recuserve h}
h((N-n)\dt, N\dt)=h((N-n)\dt, (N-n+1)\dt)+h((N-n+1)\dt, N\dt).
\end{equation}
Let $d(n,N) = h((N-n)\dt, N\dt)$ and $c(n,N) = h((N-n)\dt, (N-n+1)\dt)$.
Then \eqref{eq recuserve h} is equal to $d(n,N) = c(n,N)+d(n-1,N)$.
and we obtain $d(n,N) = d(0,N)+\sum_{j=1}^nc(j,N)$.
Since $d(0,N)=h(N\dt,N\dt)= 0$, we have $d(n,N) = \sum_{j=1}^nc(j,N)$.

Let $\eta(T)$ be a random variable with support on $\{\dt,2\dt,\cdots,N\dt = T\}$.
Suppose $c(j,N)=C(T) P(\eta(T) = j\dt)$, $j=1,2,\cdots,N$, for a constant $C(t)>0$. Then we have $d(n,N) =$ $C(T)\sum_{j=1}^n$ $P(\eta(T) = j\dt)=C(T)P(\eta(T)\le n\dt = t)$ and hence
\begin{equation}\label{eq H(t,T)}
H(t,T)=\exp(h(m\dt,N\dt))=\exp(d(n,N))=\exp(C(T) P(\eta(T)\le t)).
\end{equation}
In the case when $P(\eta(T) = j\dt)=\frac{1}{N}$ for all $j\in\{1,2,\cdots,N\}$, 
we have
\[
H(t,T)=\exp\left(\frac{nC(T)}{N}\right)=\left(\exp\left(\frac{C(T)}{N\dt}\right)\right)^{n\dt}=\left(\exp\left(-\frac{C(T)}{T}\right)\right)^{-(T-t)}.
\]
By setting $\delta=\exp\left(-\frac{C(T)}{T}\right)$, we obtain $H(t,T)=\delta^{-(T-t)}$, which gives the special case of the original Ho-Lee model.

\section{Risk-Neutral Dynamics of the Ho-Lee Binomial Interest Rate Model with Time-Dependent Parameters}
We now turn our attention to the risk-neutral dynamics implied by the binomial tree. In \cite{HoLee:1986} a portfolio of two bonds with different maturities, $S>0$ and $T>0$, is considered. This leads to the no-arbitrage condition: for some $q\in(0,1)$,
\begin{equation}\label{eq NoArbVond}
qU(\varepsilon(t),t,T)+(1-q)D(\varepsilon(t),t,T) = 1.
\end{equation}

In what follows we will choose $S=T+\dt>0$, which will be a less stringent no-arbitrage condition than \eqref{eq NoArbVond}. To this end, consider a portfolio $\pi$ consisting of (a) one unit of the zero-coupon bond with maturity $T$, and (b) $b$ units of a zero-coupon bond with maturity $S$. At time $t-\dt\ge0$, the portfolio value, denoted by $V(t-dt,T,T+\dt|E(t-\dt))$, is given by
\[  
V(t-dt,T,T+\dt|E(t-\dt))=B(t-dt,T|E(t-\dt))+bB(t-dt,T+\dt|E(t-\dt)). 
\]
In the next period, $t$, the portfolio value is given by
\begin{align*}
V(t,T,T+\dt|E(t))&=B(t,T|E(t))+bB(t,T+\dt|E(t)) \\
&= \begin{cases}
 V^U(t,T,T+\dt) &\text{ if } \varepsilon(t) = 1\\
 V^D(t,T,T+\dt)  &\text{ if } \varepsilon(t) = 0
\end{cases},
\end{align*}
where
\begin{align*}
V^U(t,T,T+\dt) =\frac{B(t-\dt,T|E(t-\dt))}{B(t-\dt,t|E(t-\dt))}U(t, T)+b \frac{B(t-\dt,T+\dt|E(t-\dt))}{B(t-\dt,t|E(t-\dt))}U(t, T+\dt)
\end{align*}
and
\begin{align*}
V^D(t,T,T+\dt) =\frac{B(t-\dt,T|E(t-\dt))}{B(t-\dt,t|E(t-\dt))}D(t, T)+b \frac{B(t-\dt,T+\dt|E(t-\dt))}{B(t-\dt,t|E(t-\dt))}D(t, T+\dt)
\end{align*}
We choose $b$ so that the portfolio becomes riskless in $[t,t+\dt]$; that is
\[
V^U(t,T,T+\dt|E(t)) =V^D(t,T,T+\dt|E(t)). 
\]
Then
\begin{equation}\label{eq b}
b = \frac{B(t-\dt,T|E(t-\dt))(D(t,T)-U(t,T))}{B(t-\dt,T+\dt|E(t-\dt))(U(t,T+\dt)-D(t,T+\dt))}.
\end{equation}
To avoid arbitrage opportunities,
\begin{align*}
V(t-\dt,T,T+\dt|E(t-\dt))
&=B(t-\dt,t|E(t-\dt))V(t,T,T+\dt|E(t)).
\end{align*}
Hence,
\begin{align}
\nonumber &B(t-\dt,T|E(t-\dt))+bB(t-\dt,T+\dt|E(t-\dt))\\
\label{eq noarb1} &=B(t-\dt,T|E(t-\dt))U(t, T+\dt)+b B(t-\dt,T+\dt|E(t-\dt))U(t, T+\dt).
\end{align}

Now \eqref{eq b} and \eqref{eq noarb1} imply
\begin{align}
\nonumber &1+\frac{D(t,T)-U(t,T)}{U(t,T+\dt)-D(t,T+\dt)}\\
\nonumber &=U(t, T+\dt)+\frac{D(t,T)-U(t,T)}{U(t,T+\dt)-D(t,T+\dt)}U(t, T+\dt)\\
\label{eq noarb2}
\text{or }~~~&\frac{U(t,T+\dt)}{U(t,T)-D(t,T)}(1-D(t,T))
-\frac{D(t,T+\dt)}{U(t,T)-D(t,T)}(1-U(t, T))=1.
\end{align}
Assuming that $U(t,T)$ and $D(t,T)$ have continuous derivatives $\frac{\partial U(t,T)}{\partial T}$ and  $\frac{\partial D(t,T)}{\partial T}$, respectively, equation \eqref{eq noarb2} implies
\[
\frac{\partial U(t,T)}{\partial T}(1-D(t,T))
-\frac{\partial D(t,T)}{\partial T}(1-U(t, T))=0.
\]
Thus, there is a positive function $\lambda(t)$ such that $\ln(U(t,T)-1)=\ln(1-D(t,T))+\ln \lambda(t)$.
Let $q(t) = \frac{1}{1+\lambda(t)}\in(0,1)$. This leads relaxed no-arbitrage condition \eqref{eq NoArbVond}:
\begin{equation}\label{eq NoArbVond2}
q(t)U(\varepsilon(t),t,T)+(1-q(t))D(\varepsilon(t),t,T) = 1.
\end{equation}
While we do not require that $q(t)=q$ for all $t\ge0$, we do require that $q(t)\in(0,t)$ for $0\le t\le T$ and $t\in\Pi(\dt)$.

Next, from \eqref{eq H(t,T)}, we have
$
\frac{U(t,T)}{D(t,T)}=H(t,T)=\exp(C(T)P(\eta(T)\le t)),
$
and hence
\[
U(t,T)=D(t,T)\exp(C(T)P(\eta(T)\le t)).
\]
By \eqref{eq NoArbVond2}, we obtain
\begin{equation}\label{eq Up}
U(t,T)=\frac{1}{q(t)+(1-q(t))\exp(-C(T)P(\eta(T)\le t))}>1, 
\end{equation}
and
\begin{equation}\label{eq Down}
D(t,T)=\frac{\exp(-C(T)P(\eta(T)\le t))}{q(t)+(1-q(t))\exp(-C(T)P(\eta(T)\le t))}<1.
\end{equation}
In the case $q(t)=q$ and $P(\eta(T)=j\dt)=\frac{1}{N}$,
we obtain Ho-Lee expressions:
\begin{equation}\label{eq UpDown HoLee}
U(t,T)=\frac{1}{q+(1-q)\delta^{-(T-t)}},~~~ \text{ and } ~~~D(t,T)=\frac{\delta^{-(T-t)}}{q+(1-q)\delta^{-(T-t)}},
\end{equation}
where $\delta=\exp(-C(T)/T)$.
\section{A Resolution of Ho-Lee Model's Shortcoming}
Let the riskless return in the one-step time-interval $(t,t+\dt]$ be $r(t,t+\dt|E(t))$. Then, to avoid arbitrage, we have $ B(t,t+\dt|E(t))\exp(r(t,t+\dt|E(t)))=1$. 
By \eqref{eq BinTree}, we have 
\[
B(t, t+\dt, E(t))=\frac{B(t,T|E(t))}{B(t+\dt,T|E(t+\dt))}\mathbb{H}(t+\dt,T).
\]
Suppose $\lim_{T\uparrow\infty}\frac{B(t+\dt,T|E(t+\dt))}{B(t,T|E(t))}=1+f^\infty(t,t+\dt|E(t+\dt))$ is nonzero and exists. Let
\[
F^{\infty}(t,t+\dt|E(t+\dt))=\frac{1}{1+f^\infty(t,t+\dt|E(t+\dt))}>0.
\]

\cite{BlissRonn:1989} pointed out the following weakness of the Ho-Lee model. 
Assume that $T\uparrow\infty$ and after a fixed moment $t\ge0$ only downturn moves had happened and $1+f^\infty(t,t+\dt|E(t+\dt))<\frac{1}{1-q}$.
Then, from \eqref{eq UpDown HoLee}, it follows that 
\begin{align}
\nonumber
\exp(-r(t,t+\dt|E(t)))&=B(t, t+\dt, E(t))\\
\nonumber
&=\lim_{T\uparrow\infty}\frac{B(t,T|E(t))}{B(t+\dt,T|E(t+\dt))}D(t+\dt,T)\\
\nonumber
&=F^{\infty}(t,t+\dt|E(t+\dt))\lim_{T\uparrow\infty}\frac{\delta^{-(T-t-\dt)}}{q+(1-q)\delta^{-(T-t-\dt)}}\\
\label{eq HoLee Weak 1}
&=\frac{F^{\infty}(t,t+\dt|E(t+\dt))}{1-q}>1,
\end{align}
resulting in negative short-term interest rates  $r(t,t+\dt|E(t))<0$.
Suppose next that after a fixed moment $t\ge0$, only upturn moves occur. Then, from \eqref{eq UpDown HoLee}, we have
\begin{align}
\nonumber
\exp(-r(t,t+\dt|E(t)))&=B(t, t+\dt, E(t))\\
\nonumber
&=\lim_{T\uparrow\infty}\frac{B(t,T|E(t))}{B(t+\dt,T|E(t+\dt))}U(t+\dt,T)\\
\label{eq HoLee Weak 2}
&=F^{\infty}(t,t+\dt|E(t+\dt))\lim_{T\uparrow\infty}\frac{1}{q+(1-q)\delta^{-(T-t-\dt)}}=0,
\end{align}
which implies $r(t,t+\dt|E(t))=\infty$, meaning that the one-period short rate at $t\ge0$ will be unbounded if a sufficiently long sequence of downturns is observed. 

Suppose \eqref{eq Up} and \eqref{eq Down}. Assume that $\eta(T)$ is uniformly distributed on $t\in\{\dt, 2\dt, \cdots, N\dt = T\}$. Then $P(\eta(T)\le t)=t/T$ and hence 
$\lim_{T\uparrow\infty} \exp(C(T)P(\eta(T)\le t))=1$.
Thus, instead of \eqref{eq HoLee Weak 1}, we have
\begin{align}
\nonumber
\exp(-r(t,t+\dt|E(t)))&=B(t, t+\dt, E(t))\\
\nonumber&=F^{\infty}(t,t+\dt|E(t+\dt))\lim_{T\uparrow\infty}\frac{\exp(C(T)P(\eta(T)\le t))}{q(t)+(1-q(t))\exp(C(T)P(\eta(T)\le t))}\\
\label{eq genHoLee1}
&=F^{\infty}(t,t+\dt|E(t+\dt))<1,
\end{align}
implying that the short rate $r(t,t+\dt|E(t))$ is positive.
Similarly, instead of \eqref{eq HoLee Weak 2}, we have
\begin{align}
\nonumber
\exp(-r(t,t+\dt|E(t)))&=B(t, t+\dt, E(t))\\
\nonumber
&=F^{\infty}(t,t+\dt|E(t+\dt))\lim_{T\uparrow\infty}\frac{1}{q(t)+(1-q(t))\exp(C(T)P(\eta(T)\le t))}\\
\label{eq genHoLee2}
&=F^{\infty}(t,t+\dt|E(t+\dt))>0,
\end{align}
implying that the short rate $r(t,t+\dt|E(t))$ is finite.Thus, the issue with respect to the drawback of the Ho-Lee model described in \eqref{eq HoLee Weak 1} and \eqref{eq HoLee Weak 2} is now resolved due to \eqref{eq genHoLee1} and \eqref{eq genHoLee2}. 

\section{Conclusion}
We extend the classical Ho-Lee binomial pricing model to the case when the model parameters are time-dependent and resolve a shortcoming of the model. We achieve this by introducing a new more flexible no-arbitrage condition, leading to a more realistic and flexible expressions for the up and down movements in the binomial pricing tree.
\clearpage

\begin{center}

\begin{footnotesize}
\begin{tabular}{llllll}
  & & & &$B^{UU}(t+2\dt,T|E(t+2\dt))$ \\
  & & & &$=\frac{B^U(t+\dt, T|E(t+\dt))}{B^U(t+\dt, t+2\dt|E(t+\dt))} U(t+2\dt, T)$\\
  & & &$\nearrow$  \\
  & & $B^U(t+\dt,T|E(t+\dt))$& \\
  & &$=\frac{B(t, T|E(t))}{B(t, t+\dt|E(t))} U(t+\dt, T)$\\
  & $\nearrow$ & & $\searrow$&\\
  & & & &$B^{UD}(t+2\dt,T|E(t+2\dt))$ \\
  & & & &$=\frac{B^U(t+\dt, T|E(t+\dt))}{B^U(t+\dt, t+2\dt|E(t+\dt))} D(t+2\dt, T)$\\
 $B(t,T|E(t))$ & & & & & \\
  & & & &$B^{DU}(t+2\dt,T|E(t+2\dt))$ \\
  & & & &$=\frac{B^D(t+\dt, T|E(t+\dt))}{B^D(t+\dt, t+2\dt|E(t+\dt))}  U(t+2\dt, T)$\\
  & $\searrow$ & &  $\nearrow$ &\\
  & & $B^D(t+\dt,T|E(t+\dt))$&  &\\
  & &$=\frac{B(t, T|E(t))}{B(t, t+\dt|E(t))} D(t+\dt, T)$\\
  & & & $\searrow$ \\
  & & & &$B^{DD}(t+2\dt,T|E(t+2\dt))$ \\
  & & & &$=\frac{B^D(t+\dt, T|E(t+\dt))}{B^D(t+\dt, t+2\dt|E(t+\dt))} D(t+2\dt, T)$\\
\end{tabular}
\end{footnotesize}
\\
~\\
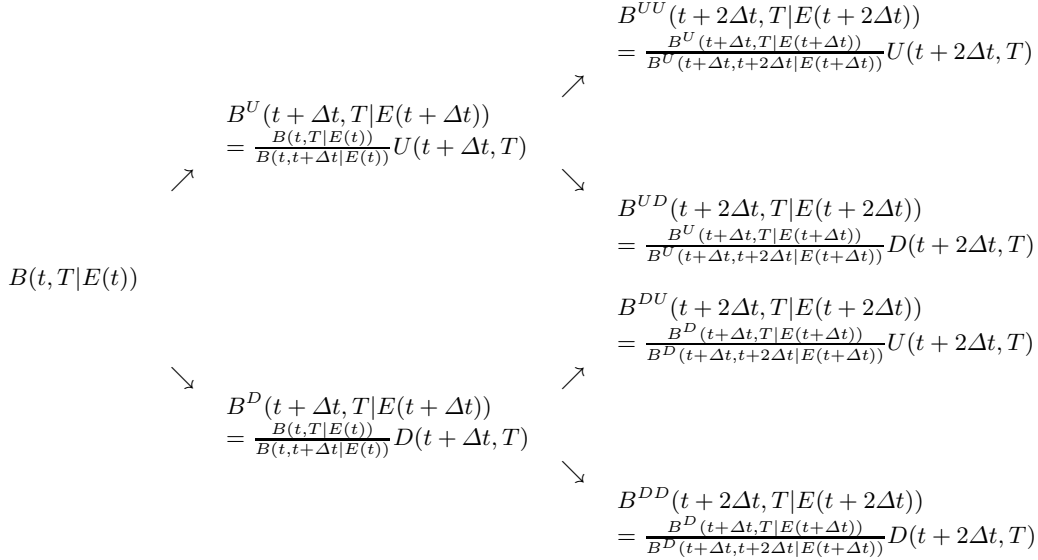
\textbf{Figure A:} Two step binomial tree model for zero price
\end{center}

\section*{Appendix}
Figure A 
 shows a two-step binomial tree based on equation \eqref{eq BinTree} at time $t\in\Pi(\dt)$.
Applying $T = t+2\dt$ to the binomial tree, we have $
B^U(t+\dt,t+2\dt|E(t+\dt))=\frac{B(t, t+2\dt|E(t))}{B(t, t+\dt|E(t))} U(t+\dt, t+2\dt),
$
and
$
B^D(t+\dt,t+2\dt|E(t+\dt))=\frac{B(t, t+2\dt|E(t))}{B(t, t+\dt|E(t))} D(t+\dt, t+2\dt)
$. 
In order to have a recombined tree, 
we put
$
B^{UD}(t+2\dt,T|E(t+2\dt)) = B^{DU}(t+2\dt,T|E(t+2\dt))
$ 
or
\begin{align*}
&\frac{\frac{B(t, T|E(t))}{B(t, t+\dt|E(t))}}{\frac{B(t, t+2\dt|E(t))}{B(t, t+\dt|E(t))} U(t+\dt, t+2\dt)} U(t+\dt, T)D(t+2\dt, T)\\
&=\frac{\frac{B(t, T|E(t))}{B(t, t+\dt|E(t))}}{\frac{B(t, t+2\dt|E(t))}{B(t, t+\dt|E(t))} D(t+\dt, t+2\dt)} D(t+\dt, T)U(t+2\dt, T).
\end{align*}
We then obtain
\[
\frac{U(t+\dt)}{D(t+\dt)} = \frac{U(t+\dt, t+2\dt)U(t+2\dt)}{D(t+\dt, t+2\dt)D(t+2\dt)}.
\]

\singlespace
\bibliographystyle{decsci_mod}
\bibliography{refs_HooLeeTree}

\end{document}